**Title:**

# Correlative Theoretical and Experimental Study of the Polycarbonate | X Interfacial Bond Formation (X = AlN, TiN, TiAlN) during Magnetron Sputtering

**Authors:**

*Lena Patterer\*, Pavel Ondračka, Dimitri Bogdanovski, Stanislav Mráz, Soheil Karimi Aghda, Peter J. Pöllmann, Yu-Ping Chien, Jochen M. Schneider*

L. Patterer*

Materials Chemistry, RWTH Aachen University, Kopernikusstr. 10, 52074 Aachen, Germany

*Corresponding author: patterer@mch.rwth-aachen.de

Dr. P. Ondračka

Department of Physical Electronics, Faculty of Science, Masaryk University, Kotlářská 2, 611 37 Brno, Czech Republic

Dr. D. Bogdanovski, Dr. S. Mráz, S. Karimi Aghda, P.J. Pöllmann, Y.-P. Chien, Prof. Dr. J.M. Schneider

Materials Chemistry, RWTH Aachen University, Kopernikusstr. 10, 52074 Aachen, Germany

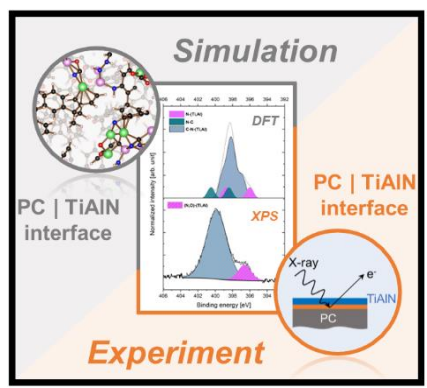

**Abstract**

To understand the interfacial bond formation between polycarbonate (PC) and magnetron-sputtered metal nitride thin films, PC | X interfaces (X = AlN, TiN, TiAlN) are comparatively investigated by *ab initio* simulations as well as X-ray photoelectron spectroscopy. The simulations predict significant differences at the interface, as N and Ti form bonds with all functional groups of the polymer, while Al reacts selectively only with the carbonate group of pristine PC. In good agreement with simulations, experimental data reveal that the PC | AlN and the PC | TiAlN interfaces are mainly defined by interfacial C-N bonds, whereas for PC | TiN, the interface formation is also characterized by numerous C-Ti and (C-O)-Ti bonds. Bond strength calculations combined with the measured interfacial bond density indicate the strongest interface for PC | TiAlN followed by PC | AlN, whereas the weakest is predicted for PC | TiN due to its lower density of strong interfacial C-N bonds. This study shows that the employed computational strategy enables prediction of the interfacial bond formation between PC and metal nitrides and that it is reasonable to assume that the research strategy proposed herein can be readily adapted to other organic | inorganic interfaces.

1. **Introduction**

Poly(bisphenol A carbonate), often referred to as polycarbonate (PC), is a widely used thermoplastic polymer in the plastics industry. PC and its blends are typically used in data-recording media (e.g. DVDs),[1,2] medical equipment (e.g. blood reservoirs),[1,2] computer and printer housing,[1,2] automobile interior (e.g. instrument panels, consoles), as well as exterior parts (e.g. automotive bumpers).[1–5] However, surface coatings are often needed since PC is prone to stress corrosion cracking, is sensitive to scratches, and exhibits poor resistance to commonly used industrial solvents and automotive fluids.[2]

Cubic TiAlN (space group $Fm\bar{3}m$) is a commonly applied protective thin film for cutting and forming tools. Its outstanding oxidation,[6,7] corrosion,[8] and wear resistance[6] make TiAlN also an attractive candidate for the protection of polymer components, even though literature on this subject is scarce. To fulfill the purpose of a protective thin film, a strongly adhering interface with the substrate material is required. As stresses would adversely affect adhesion, magnetron sputtering with floating substrate potential,[9,10] no intentional heating,[10] as well as a low film thickness,[10] are employed in this study to maximize the interfacial strength. However, besides the interfacial stress, the chemical bonding between the thin film and the polymer is of significance to validate the adhesion properties[11] and is the focus of this work.

The chemical structure of PC consists of hydrocarbon aromatic and aliphatic groups as well as the carbonate group (O-(C=O)-O),[12] which are all possible interaction sites for the interface formation. Several studies explored the interfacial bond formation between PC and different metals, like Au,[13] Al,[14] or Cr[15]. For the precious metal Au, the C=O bond of the carbonate group was identified as the primary interaction site, while only a weak interaction with the aromatic groups was detected.[13] Similarly, it was shown that the PC | Al interface is mainly defined by reactions with the carbonate group and diffusion of Al atoms into the polymer.[14] After

evaporating the transition metal Cr onto PC, however, simultaneous bond formation with both the carbonate and aromatic groups was shown to occur.[15]

When considering the comparably complex polymer | ceramic interfaces in literature, Pedrosa et al.[16] investigated the chemically-induced changes in the adhesion behavior of TiN$_x$ thin films deposited onto PC substrates for bio-electrode applications and showed that the films close to the 1:1 stoichiometry ($x = 0.95$) exhibit significantly better adhesion properties compared to films consisting of α-Ti with interstitially incorporated N ($x = 0.24$). In another study, the chemical bonding between plasma-treated polyethylene (PE) and the subsequently deposited Ag was explored.[17] Gerenser[17] demonstrated that the adhesion strength of Ag depends on the chosen working gas for the plasma pre-treatment in the following order: untreated < argon < oxygen < nitrogen.[17] For the nitrogen-plasma treated PE | Ag interface, (C-N)-Ag bonds were identified by X-ray photoelectron spectroscopy (XPS), increasing the interfacial adhesion strength (peel force) of the Ag thin film by a factor of 8 compared to the untreated sample.[17] Furthermore, Klemberg-Sapieha et al.[18] investigated the PC | SiN$_{1.3}$ interface by XPS and reported the formation of C-Si, (C-O)-Si, and C-N-Si bonds. Additionally, they evaluated the influence of different polymer pre-treatments on the adhesion of the SiN$_{1.3}$ film determined by scratch tests, also identifying nitrogen as the most effective and hydrogen as the least effective working gas. Similar observations were reported by Chen et al.[19], where SiN$_x$ was deposited onto nitrogen-plasma pre-treated PC. All these studies indicate that N plays a crucial role in the formation of strongly adhering polymer | thin film interfaces.

Utilizing theoretical approaches to evaluate the interfacial bond formation of thin films with polymers, many studies focused on placing individual atoms at possible reaction sites of the polymers using *ab initio*-based calculations. This approach was recently used, for instance, to investigate the metalization of poly-epoxy with Cu[20] and Al[21]. While these small systems have the advantage of providing reasonable computational speed, they are a strongly simplified representation of the final state after deposition, whereas modeling the actual processes happening during the deposition is neglected. However, some studies use classical molecular dynamics (CMD) instead to examine interface reactions with polymers. Recently, the interface of Al mechanically joined with polypropylene (PP) was simulated by CMD using an interface consisting of 4000 atoms for PP and 11000 atoms for Al.[22] Even after annealing the system at 1000 K, no covalent C-Al bonds were identified at the interface.[22] Also, for exploring the effect of kinetic energy and cluster size of sputtered $Ar_n$ clusters onto PE and polystyrene (PS), consisting of a box with $< 10^5$ atoms, CMD was used.[23] As another example, CMD was applied to investigate the film growth of TiN as a function of the kinetic energy of bombarding species.[24] From these studies, it is obvious that most film growth simulations are carried out using CMD since larger systems and longer time scales can be considered due to the significantly lower computational costs. However, *ab initio* molecular dynamics (AIMD) simulations have the advantage of higher accuracy and their application is not limited to the availability of suitable potentials which are required for CMD to define the interatomic forces.[25]

The lack of computational studies in literature on the mechanism of the interface formation between sputtered metal nitrides and PC - especially by using the accurate method of AIMD - and also the scarcity of experimental data for a systematic investigation of the bond formation between PC and $Ti_xAl_{1-x}N$ motivated the research communicated herein. Therefore, we systematically examined reactively sputtered AlN, TiN, and TiAlN thin films deposited onto PC in a correlative theoretical and experimental approach to identify the ideal chemical composition for an adhering interface.

## 2. Method development

### 2.1. Characterization of deposited thin films

The following metal-to-nitrogen ratios in terms of atomic composition were measured by energy dispersive X-ray spectroscopy for thin films (thickness > 100 nm) deposited onto Si substrates: $Al_{0.51}N_{0.49}$, $Ti_{0.50}N_{0.50}$, and $Ti_{0.20}Al_{0.27}N_{0.53}$. Additionally, these samples were used to determine the film thickness and the nominal deposition rate by analyzing their cross-sections. Deposition rates of 0.17, 0.19, and 0.23 nm s$^{-1}$ were determined for the AlN, TiN, and TiAlN deposition series performed in this study, respectively. Since the probing depth of XPS is only 5-10 nm,[26] a film thickness of ~ 1 nm was aimed for the PC | X interface samples (X = AlN, TiN, TiAlN) to obtain a sufficient signal from the interface and the underlying PC. The samples used for analyzing the C 1s and the N 1s signals of the PC | AlN, and PC | TiAlN interfaces had a nominal metal nitride film thickness of 0.85, and 1.15 nm, respectively. However, for the TiN system, two PC | TiN samples with different film thicknesses were considered in this study: For the C 1s analysis, a nominal thickness of ~ 1.14 nm of TiN was used (comparable to the AlN and TiAlN systems), whereas a thinner sample (~ 0.38 nm) was considered for the analysis of the N 1s components. At that early stage of the TiN film growth, more N 1s contributions are observed compared to the spectra after longer deposition times, indicating that the initially formed C-N components react with Ti atoms to form C-N-Ti components (see **Figure S2** in the supporting information).

### 2.2. Development of an amorphous PC surface model by AIMD

For developing a PC bulk model for the subsequent simulation of the surface bombardment, three PC chains were compressed until the typical PC density of 1.2 g cm$^{-3}$ was reached[27] (supporting information, **Figure S1**). After relaxation, a surface with H-terminated functional groups of PC was created (see plan-view of the surface in **Figure 2a**). From the pristine PC surface model, the theoretical binding energies (BEs) of the functional groups were calculated and compared to the XPS spectrum of spin-coated PC (**Figure 1**). Both spectra reveal the existence of the hydrocarbon component (C-C, C-H, C=C, blue), the $C_{ring}$-O component (orange), and the carbonate group (O-(C=O)-O, pink). For the experimental C 1s spectrum, also the π-π* shake-up satellite signals, associated with aromatic bonding, at BE = 291.5 and 292.5 eV are detected, which is in accordance with literature.[28] Very good agreement between the theoretical and experimental BEs is determined for the hydrocarbon group at BE = 284.6 and the $C_{ring}$-O group at BE = 286.1 eV (**Figure 1a** & **b**). However, the BE of the carbonate group is slightly underestimated by DFT (BE = 289.3 eV) compared to the experiment (BE = 290.4 eV). While errors on the order of 1 eV compared to the experimental values are expected,[29,30] the trend of the chemical shift determined theoretically is in good agreement with the experimental spectrum.

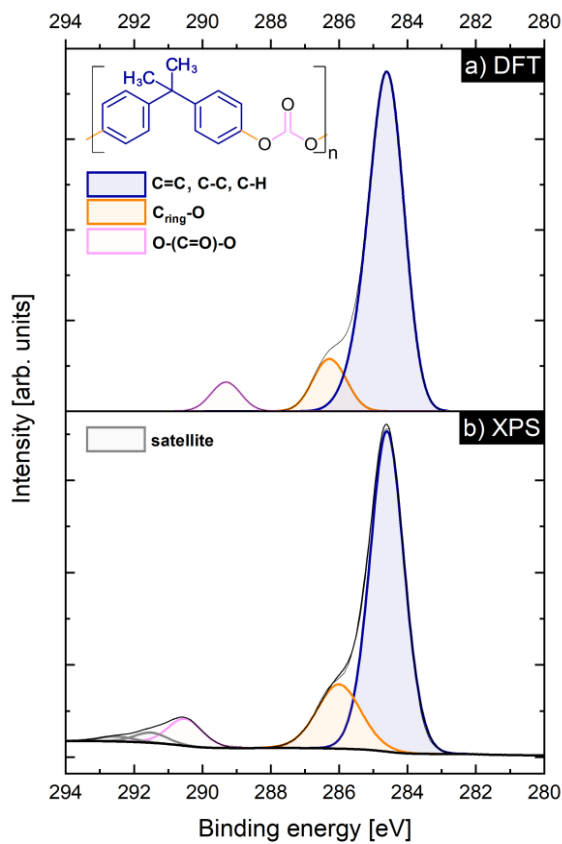

**Figure 1.** C 1s spectra of a) the simulated PC bulk model calculated by DFT, and b) spin-coated PC measured by XPS.

## 3. Results and discussion

### 3.1. Sputter deposition simulation

For each sputter deposition simulation of metal nitride onto PC, the bond formation at the interface is investigated after the deposition of 30 atoms (plan view of interfaces shown in **Figure 2b-d**). Comparing the different nitrides, the simulations reveal significant differences regarding the reactivity of the atoms and the resulting interface formation.

In **Figure 2b**, the PC|AlN interface bonding after deposition is depicted. Cluster formation of the deposited Al atoms is observed, as most interfacial bonds are defined by C-N bonds, while Al reacts mainly with C-N and C-O groups by the formation of N-Al or O-Al bonds. Only a few bonds are visible between Al and C atoms, explainable by the low reactivity between

Al atoms and hydrocarbon groups observed previously.[14,22,31] Analysis of the MD trajectories of the PC | AlN simulation[32] reveals that Al atoms are repelled by the hydrocarbon groups. Reactions between Al and (hydro-)carbon groups are only observed if C radicals were created due to previous bombardment events. In this way, mostly C-N bonds and some C-Al and (C-O)-Al bonds are visible at the PC | AlN interface, while Al preferentially clusters around C-N groups (**Figure 2b**).

After the TiN sputter deposition simulations, the formation of various C-Ti, (C-O)-Ti, and C-N groups is observed at the interface (**Figure 2c**) indicating a high reactivity of both Ti and N with PC. Overall, the interface exhibits a significantly higher bond density compared to the PC | AlN interface (**Figure 2b**). Additionally, O atoms from the carbonate group of PC seem to be highly attracted by Ti atoms, forming complex interfacial groups consisting of C, O, Ti, and N atoms (**Figure 2c**).

The highest degree of intermixing atoms within the interface layer is observed for the simulated PC | TiAlN interface, leading to a cross-linking of complex (C,O)-(Al,Ti,N) groups (**Figure 2d**). It seems that the interface formation is characterized by many bond-breaking events due to the high reactivity of Ti and N atoms, resulting in a PC surface with many reactive groups (radicals). In this way, also Al contributes to a higher extent to the interfacial bond formation compared to the simulated PC | AlN interface (compare **Figure 2b** & **d**).

Comparing the three interfaces after the simulated metal nitride depositions, the PC | AlN interface is mainly defined by the formation of C-N groups surrounded by Al clusters, while a significantly higher PC surface coverage is observed after the deposition of TiAlN, leading to a high degree of cross-linking and intermixing of complex interfacial groups for TiAlN. Also, the PC | TiN interface shows more interfacial bonds compared to the PC | AlN interface and is characterized by both N- and Ti-based interfacial groups. The exact quantification of interfacial bonds will be discussed during the integrated crystal orbital Hamilton population analysis (ICOHP) analysis in chapter 3.3.

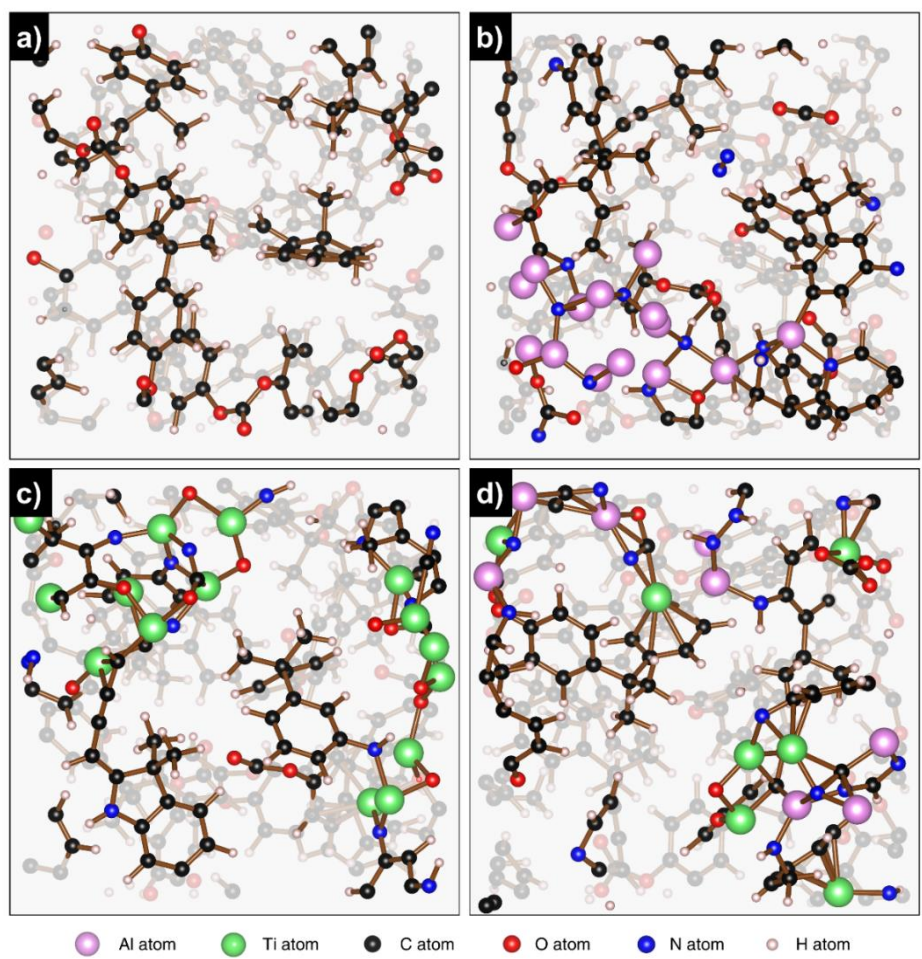

**Figure 2.** Plan view of the PC surface slab model for a) the pristine surface and the surfaces after the deposition of 30 atoms of b) AlN, c) TiN, as well as d) TiAlN visualized by VESTA [33].

### 3.2. Comparison of the theoretical and experimental bond formation

To compare the theoretical work with experiments, BE calculations of the simulated interfaces were performed by DFT and compared with the XPS spectra of the interfaces obtained from magnetron sputtering experiments. The focus of this work lies on the comparison of the C 1s and N 1s signals, whereas it is not possible to calculate the BEs of the Al 2p and Ti 2p components, due to not yet developed OpenMX pseudopotentials of these elements. The XPS C 1s signal is analyzed by comparing the spectra of pristine PC with the PC | X interface spectra (X = AlN, TiN, TiAlN), whereas the N 1s signal of the interfaces is compared to the N 1s spectra of the respective thin films (thickness > 100 nm) to identify the interface components. While the C 1s signal provides general information about new interfacial groups forming at the polymer, the N 1s signal provides explicit information about the interfacial N groups and the metal nitride film growing atop. After simulating the deposition of 30 atoms onto PC, several different configurations are identified, each described by discrete BEs of C and N atoms, which contribute to the broadening of the C 1s and N 1s spectra. To conduct a well-structured and comprehensible interface analysis, the contributions of individual atoms were categorized into different groups by adding up their relative intensities. Thus, the following interfacial groups were summarized for the theoretical C 1s spectra: (C-O)-(Al,Ti,N) groups resulting from the reaction of incident atoms with the carbonate groups, while C-Al, C-Ti, and C-N groups are characterized by the bond formation of Al, Ti, and N with the (hydro-)carbon groups, respectively.

Starting with the simulated PC | AlN interface (**Figure 3a**), the calculated C 1s spectrum reveals mainly C-N groups at the interface (teal component at BE = 285.2 eV), however, also C-Al groups are present (yellow component between BE = 284.6 – 283.3 eV). The lowest signal is observed for the (C-O)-(Al,N) component (light blue at BE = 288.6 eV), indicating the bond formation of both Al and N atoms with the carbonate group.

Experimentally, the combined component of C-N + (C-O)-(Al,N) groups at 285.5 eV (teal-blue striped component, **Figure 3b**) agrees well with the predicted position of the C-N groups. Due to the reported overlapping BE range of C-N and (C-O)-(Al,N) groups (285.9 – 288.7 eV[18,34] and 285.5 – 289.6 eV,[18,31] respectively), these groups were combined for the experimental fitting of the C 1s spectrum (**Figure 3b**). The component at 287.3 eV was also assigned to C-N + (C-O)-(Al,N) groups (**Figure 3b**), corresponding to more positively charged C atoms compared to the component at 285.5 eV. For both C-N + (C-O)-(Al,N) components, the main contribution is likely due to C-N bonds as predicted by DFT (**Figure 3a**) and also confirmed by the C-N groups detected in the N 1s spectrum (**Figure 6a**). The (C-O)-(Al,N) component at BE = 289.1 eV (light blue component, **Figure 3b**) confirms experimentally the reaction of Al and N atoms with the intact carbonate group as reported before.[18,31] For the C-Al groups, the calculated spectra predict a negative BE shift relative to the hydrocarbon group, which is verified experimentally by the detected component at BE = 282.2 eV (**Figure 3b,** Ref.[35]). Consequently, all predicted interfacial groups after AlN sputtering onto PC are confirmed experimentally by the XPS C 1s spectrum.

When comparing the relative intensities of the interfacial components, corresponding to their populations, differences are apparent as the C-Al groups have experimentally a significantly lower intensity relative to the C-N + (C-O)-(Al,N) components than for the predicted C 1s spectrum (**Figure 3b** & **a**). These differences might be explained by some limitations of the simulation model. The model assumes a stoichiometric ratio for the surface bombardment by Al and N. However, it is plausible that the first arriving atoms are characterized by a higher N concentration due to the lighter sputtered N compared to Al atoms (resulting in a faster velocity with constant kinetic energy). Combining a higher N concentration in the first layer with the selective reactivity of Al to form only bonds with C radicals rather than the intact hydrocarbon groups of PC, the low population of experimentally detected C-Al groups can be explained.

Overall, it is shown that the formation of all DFT-predicted interfacial groups for the PC | AlN interface is confirmed experimentally. Additionally, the simulation accurately demonstrated that the high reactivity of N and the selective reactivity of Al atoms are critical mechanisms for the more pronounced formation of C-N bonds compared to the C-Al bonds.

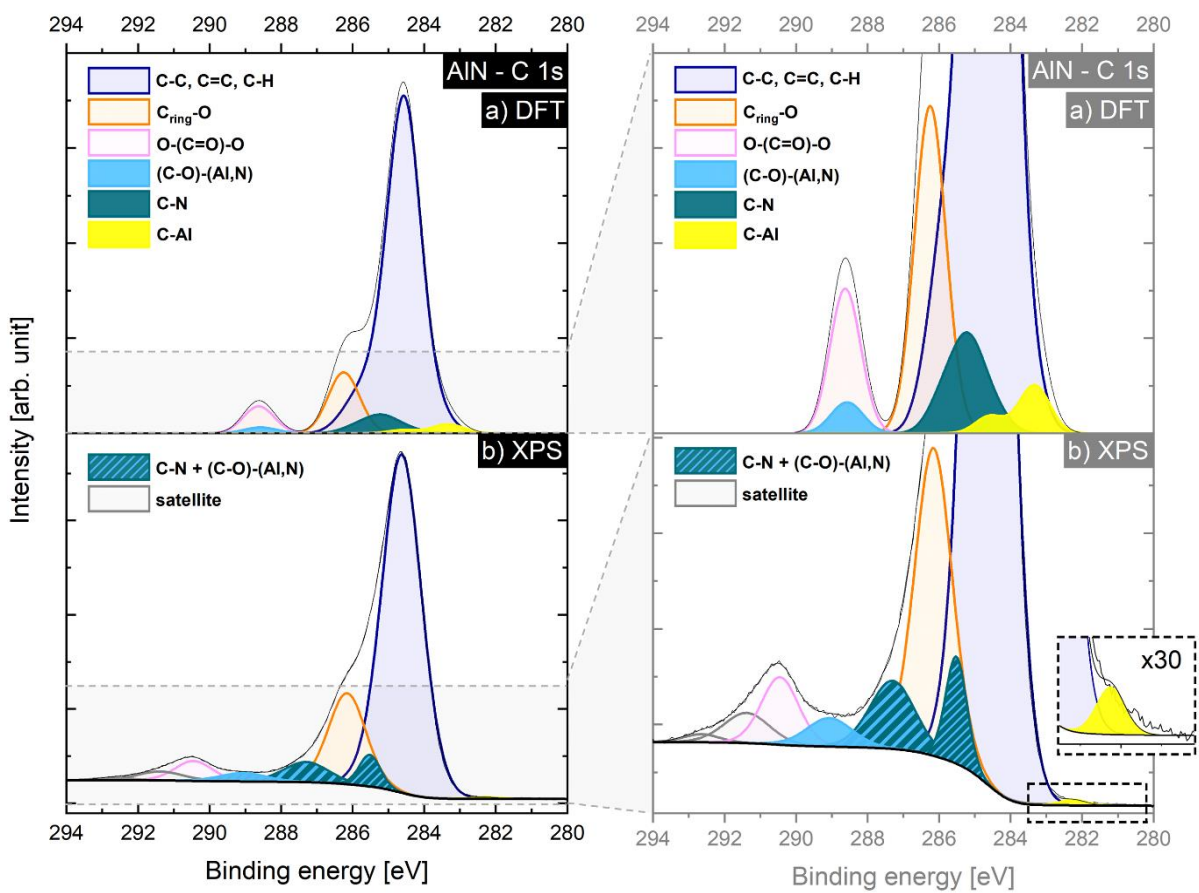

**Figure 3.** C 1s spectra a) calculated by DFT and b) measured by XPS for the PC | AlN interface shown as (left) whole spectra, and (right) magnification of the low-intensity C 1s components. The different groups are indicated by color code.

Considering the simulated PC | TiN interface (**Figure 4a**), the population of C-N and C-Ti groups is similar, as indicated by the area of the teal (BE = 285.4 eV) and green components (BE = 284.4 – 283.4 eV), respectively. Also, a (C-O)-(Ti,N) component at BE = 285.8 eV is observed in **Figure 4a** (red component), indicating reactions of deposited Ti and N atoms with the carbonate group.

The predicted interfacial C-N, C-Ti, and (C-O)-(Ti,N) components are all experimentally verified by the XPS C 1s spectrum (**Figure 4b**). The two combined components of C-N + (C-O)-(Ti,N) (teal-red striped) are observed at BEs of 285.3 and 287.6 eV, similar to the XPS C 1s spectrum of the PC | AlN interface (**Figure 3b**). The component at BE = 288.9 eV (red,

**Figure 4b**) is assigned to (C-O)-(Ti,N) groups,[18,31] while the low-BE signals at 281.9 – 283.0 eV (green, **Figure 4b**) are attributed to C-Ti bonds as reported before.[31,36]

When comparing the experimentally obtained population of the C-Ti groups, indicated by their relative intensity, at the PC | TiN interface (**Figure 4b**) with the C-Al groups at the PC | AlN interface (**Figure 3b**), it is evident that Ti is significantly more reactive towards the hydrocarbon groups of PC compared to Al as predicted by the simulations. However, considering the relative contributions in the experimental C 1s spectrum of the PC | TiN interface, the C-Ti groups constitute a lower population compared to the C-N + (C-O)-Ti groups (**Figure 4b**), whereas the simulation predicts a similar or even higher population of C-Ti groups (**Figure 4a**). This observation supports the hypothesis made previously for the AlN deposition: the concentration of N atoms of the first arriving atoms might be experimentally higher than the 50% assumed for the sputter deposition simulation, due to the lower weight and the resulting faster velocity of N among the first deposited atoms.

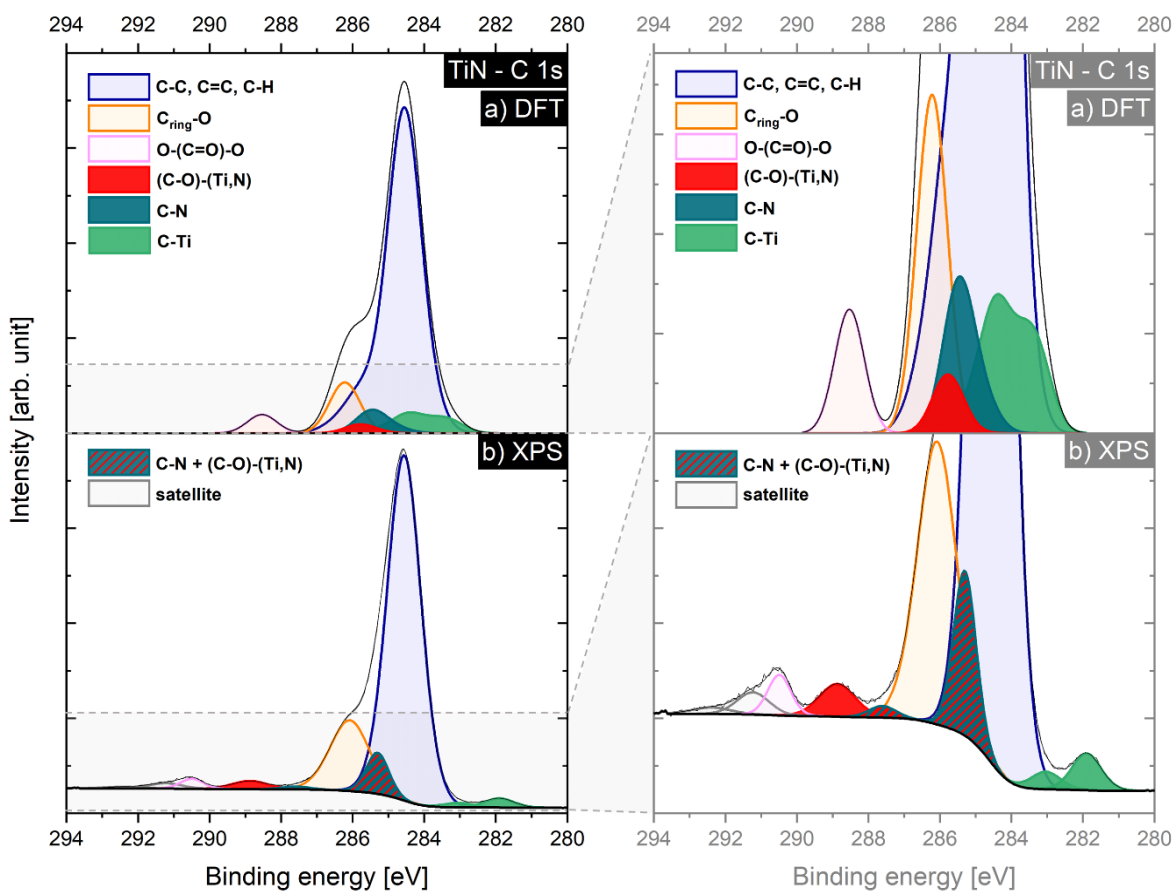

**Figure 4.** C 1s spectra a) calculated by DFT and b) measured by XPS for the PC | TiN interface shown as (left) whole spectra, and (right) magnification of the low-intensity C 1s components. The different groups are indicated by color code.

The calculated BEs of the PC | TiAlN interface constitute a similar C 1s spectrum compared to the ones of the simulated PC | AlN, and PC | TiN interfaces, as it consists of C-Ti (green), C-N (teal), C-Al (yellow), and (C-O)-(Ti,Al,N) (purple) groups (**Figure 5a**). However, the predicted population of interfacial groups is higher for the PC | TiAlN interface compared to the AlN and the TiN system as indicated by their relative intensities in the C 1s spectrum (compare **Figure 3a, Figure 4a, Figure 5a**). This higher density of interfacial bonds compared to both PC | AlN and PC | TiN interfaces reflects the complex cross-linking of interfacial groups already observed in **Figure 2d** for the simulated PC | TiAlN interface and will be quantified more precisely in the discussion of the ICOHP analysis.

When comparing the simulated interfacial bond formation (**Figure 5a**) to the XPS C 1s spectrum (**Figure 5b**), all interfacial groups are confirmed experimentally. Considering the relative intensities of the groups, again, significantly fewer C-Ti and C-Al bonds are detected experimentally (green-yellow striped component at BE = 282.3 eV, **Figure 5b**) compared to the simulation (green + yellow components, **Figure 5a**). However, large components attributed to C-N + (C-O)-(Ti,Al,N) groups are detected experimentally (teal-purple striped components at BE = 285.5 and 287.4 eV, **Figure 5b**). Assuming that the main contribution for these components is due to interfacial C-N groups (see N 1s signal, **Figure 6**), N has experimentally a significantly higher contribution to the interface formation compared to the metal atoms.

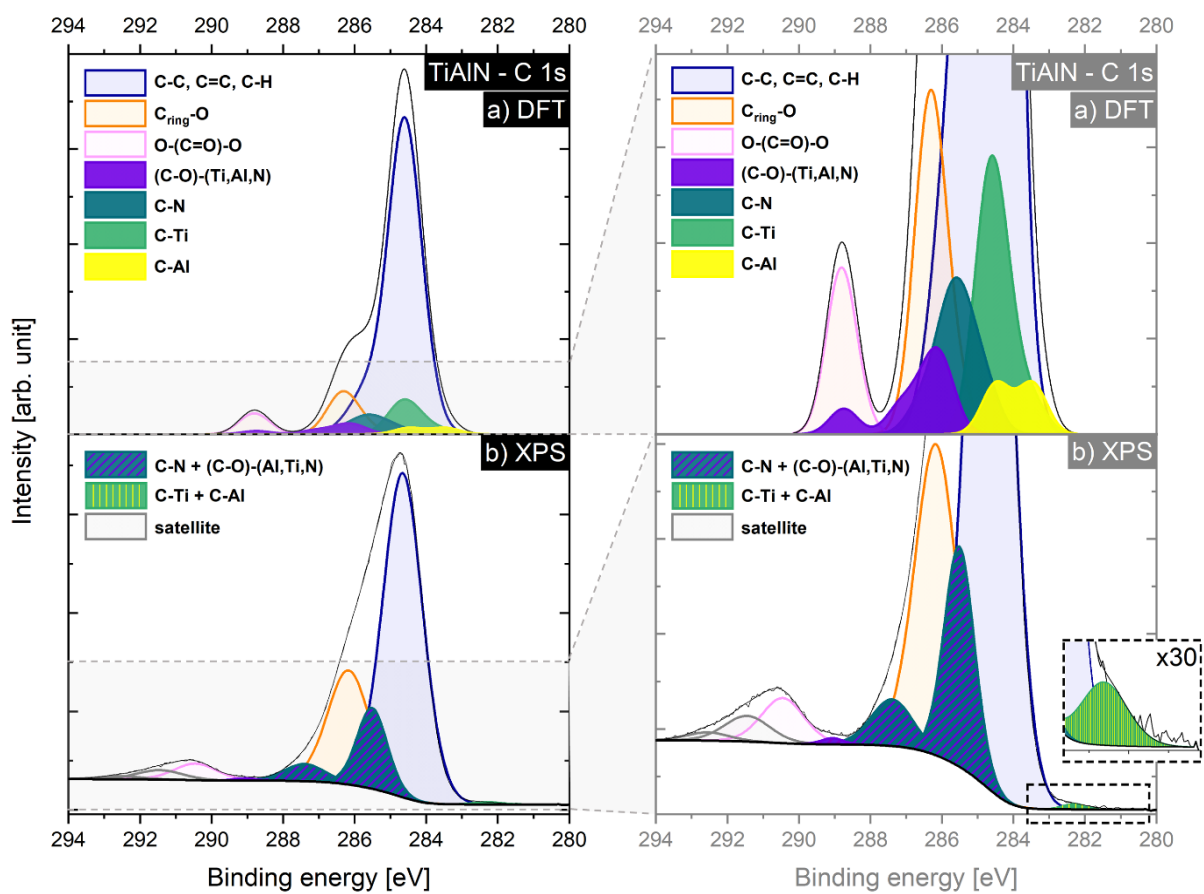

**Figure 5.** C 1s spectra a) calculated by DFT and b) measured by XPS for the PC | TiAlN interface shown as (left) whole spectra, and (right) magnification of the low-intensity C 1s components. The different groups are indicated by color code.

While the C 1s signal provides important information about the interface concerning new groups forming at the polymer, the N 1s signal provides additional information regarding the N-C groups and the thin film growth. **Figure 6** compares the calculated (top) and measured (bottom) N 1s spectra of the three interfaces. The calculated spectra are categorized mainly into three different components: N-C, C-N-metal, and N-metal groups (metal = Ti, Al). The N-C and C-N-metal groups provide information about the progress of the interface formation, while the N-metal groups belong to the forming film covering the interface.

When comparing the N 1s spectra of simulated interfaces, differences in the interface formation and the subsequent film growth are apparent: For the simulated PC | AlN interface, the population of N-C, C-N-Al, and N-Al groups are similar as indicated by their area fractions (**Figure 6a**, top). In contrast, the simulated PC | TiN interface shows a significantly higher population of complex C-N-Ti groups compared to N-C and N-Ti bonds, reflecting the high reactivity of both sputtered Ti and N atoms with PC (**Figure 6b**, top). A similar trend is observed for the PC | TiAlN interface (**Figure 6c**, top).

Experimentally, the existence of at least two N 1s components is confirmed for all interfaces. Even for the symmetric N 1s signal of the PC | AlN interface, two components are identified after considering the FWHM of the (N,O)-Al component of the respective > 100 nm-thick AlN thin film (~ 0.9 eV). The rather symmetric N 1s peak shape for PC | AlN compared to the clearly distinguishable N 1s contributions for PC | TiN can be explained by the different electronegativity of Al and Ti. The lower electronegativity of Ti compared to Al[37] leads to a higher charge transfer[38] from Ti to N compared to the charge transfer from Al to N. Thus, a higher BE shift of the N-metal compared to the N-C component is observed for the PC | TiN (ΔBE ~ 3.8 eV) compared to the PC | AlN interface (ΔBE ~ 1.2 eV), while the BE shift for the PC | TiAlN interface is in-between (ΔBE ~ 3.3 eV) (**Figure 6**, bottom). Hence, the XPS N 1s spectrum of the PC | TiN reveals four components, while the N 1s signal of the PC | AlN interface is rather symmetric. For all three interfaces, the component(s) at higher BE (> 399 eV) can be

assigned to interfacial N groups, while the lower BE signal(s) (< 399 eV) correspond to the forming metal nitride film.

The predicted and the experimental N 1s spectra of the PC | AlN interface agree well, as the presence of interfacial C-N + C-Al-N bonds and the formation of N-Al thin-film species are experimentally confirmed (**Figure 6a**). Considering the intensities in both the theoretical and experimental spectra, the interface-forming N species (N-C + C-N-Al) constitute a larger population compared to the film-forming N-Al species after the experimental deposition of ~ 1 nm AlN onto PC (**Figure 6a**).

Also, the XPS N 1s signal of the PC | TiN interface (**Figure 6b,** bottom) confirms the formation of all predicted interfacial groups: C-N (teal), C-N-Ti (teal-yellow striped), and N-Ti (yellow) groups (**Figure 6b,** top). The different area fractions of these groups determined for the theoretical model and experiment could either reflect an earlier stage of the simulated interface compared to the stage of the experimental spectrum or the differences might be caused by an experimentally different N-to-metal ratio at the beginning of the sputter deposition. The experimental N 1s spectra suggest, however, that N has a lower effect on the interface formation for the PC | TiN interface compared to the PC | AlN interface (**Figure 6b** & **a**, bottom): The C-N + C-N-Ti contributions in the N 1s signal have a lower area fraction (~ 30% with 0.38 nm TiN, and ~ 20% with 1.14 nm TiN, see **Figure S2**) compared to the fraction of the C-N + C-N-Al component (~ 70% with 0.85 nm AlN). This reflects the high reactivity of Ti leading to a similar population of C-Ti and (C-O)-Ti groups compared to C-N groups. This is supported by the experimentally detected, significantly larger C-Ti component (**Figure 4b**) compared to the C-Al component (**Figure 3b**).

Very good agreement between theory and experiment is observed for the PC | TiAlN interface as the largest signal is detected experimentally for the interface component (~ 90%, at BE = 399.9 eV,), while the population of the metal (oxy-)nitride groups is still low in intensity

(~ 10%, at BE = 396.6 eV) after the film deposition of ~ 1.15 nm. This indicates that the PC | TiAlN interface is - similar to the PC | Al interface - mainly characterized by C-N groups.

While all predicted groups for the C 1s and N 1s spectra are experimentally verified, only the population of groups differs in some cases, likely due to the experimentally higher N-to-metal ratio at the interface compared to the stoichiometric composition assumed for the AIMD simulations. Other experimental conditions that cannot be treated with the current computational protocol are e.g. the distribution in kinetic energy of the sputtered species,[39] the influence of impurities both within the polymer as well as those introduced at the interface by vapor phase condensation, and the difference in deposition rates.

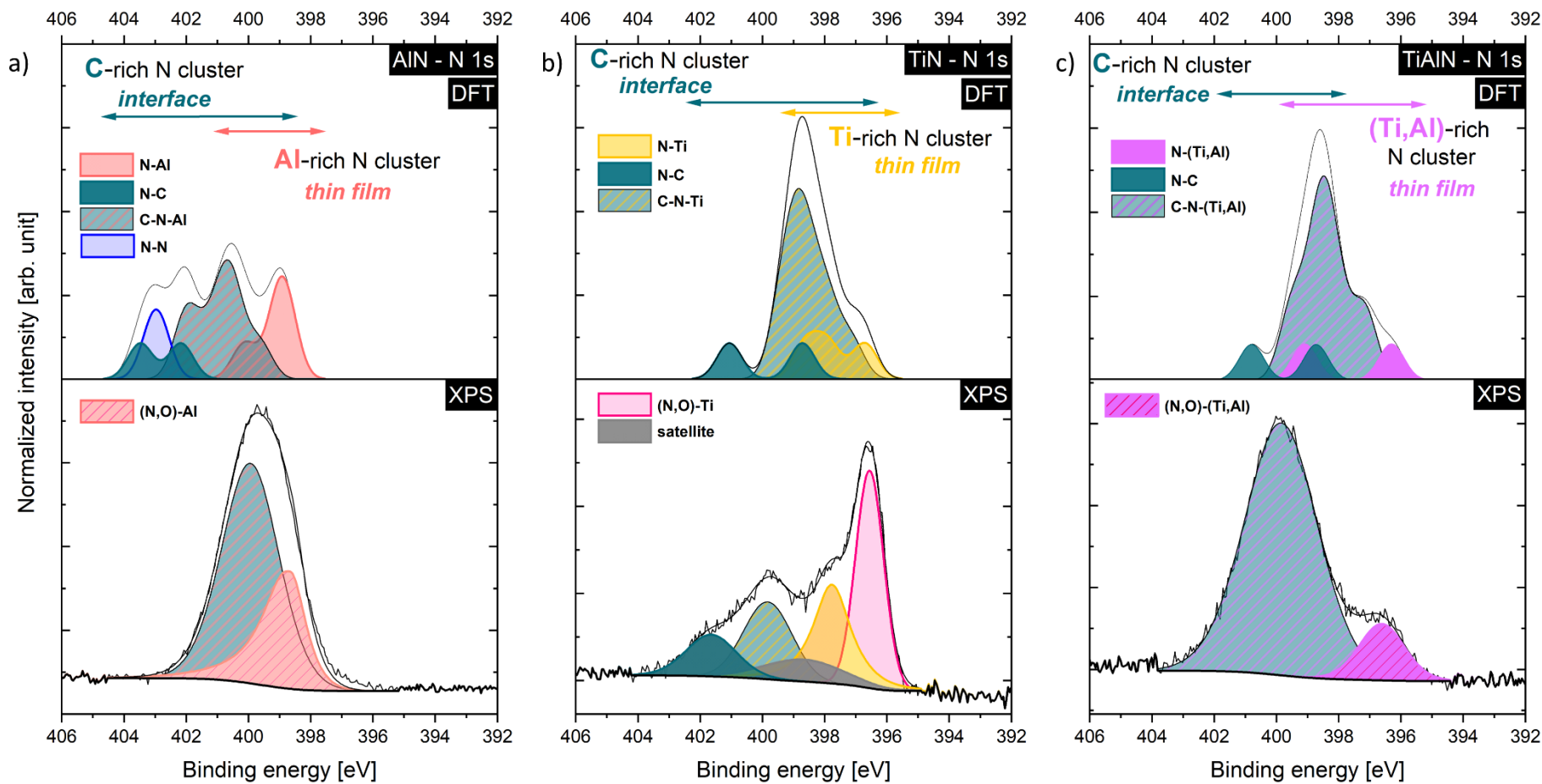

**Figure 6.** N 1s spectra of PC | X interfaces with X = a) AlN, b) TiN [40], c) TiAlN. Upper spectra are calculated by DFT, while the lower spectra are experimentally obtained by XPS. The different groups are indicated by color code.

### 3.3. ICOHP analysis of the PC | X interfaces (X = AlN, TiN, TiAlN)

To compare the strength of interfacial bonds, ICOHP analyses of the simulated PC | X interfaces (X = AlN, TiN, TiAlN) were performed. In general, the rule applies: the more negative an ICOHP value, the stronger the bond. Since the ICOHP value does not only depend on the quantum-chemical nature of the bond but also on the interatomic distance between the atoms, the ICOHP is depicted as a function of the bond length in **Figure 7**. The ICOHP values of interfacial bonds only were considered here, namely, C-N (teal), C-Al (yellow), C-Ti (green), (C-)O-Al (blue), and (C-)O-Ti (red) bonds (see **Figure 7**), where the ICOHP of (C-)O-metal bonds correspond to the bond strength between the oxygen and the metal atom, provided that the oxygen atom is still linked to a carbon atom.

As depicted in **Figure 7**, the interfacial C-N bonds – independent of the metal nitride system – exhibit the highest strength (ICOHP between −8 and −18 eV). However, the ICOHP of an average C-N bond for the PC | AlN interface is slightly lower (ICOHP = −10.7 eV) compared to the average C-N bond for PC | TiN (ICOHP = −11.6 eV), while the TiAlN system exhibits a value in-between (ICOHP = −11.1 eV). In comparison, (C-)O-Al, (C-)O-Ti, C-Al, and C-Ti bonds are significantly weaker (ICOHP ranging between −1 and −5 eV). The Al-based interfacial bonds are, however, slightly stronger compared to the Ti-based bonds as the absolute ICOHP values of the average (C-)O-Al and C-Al bonds are larger by 1.4 and 1.8 eV compared to the average (C-)O-Ti and C-Ti bonds, respectively. However, the density of C-Ti bonds (number of green data points) is predicted to be significantly higher at the PC | TiN, and PC | TiAlN interfaces (**Figure 7b** & **c**) compared to C-Al bonds (number of yellow data points) at the PC | AlN interface (**Figure 7a**).

This analysis likewise reveals the overall number of interfacial bonds forming at the simulated interfaces (**Figure 7**) and predicts the highest number of interfacial bonds for PC | TiAlN (37), followed by PC | TiN (34), while the lowest number of bonds is predicted for the

PC | AlN interface (21). These differences in the density of interfacial bonds can be explained by the high reactivity of N and Ti, while Al reacts only with reactive surface groups, such as C-N, C-O, and C radicals. While for PC | AlN, mainly cluster formation is observed, leading to a low interfacial bond density, PC | TiAlN exhibits the highest interfacial bond density due to the formation of many reactive surface groups, leading to a complex cross-linking of interfacial groups (compare **Figure 2**d). The PC | TiN interface also exhibits a higher interfacial bond density compared to PC | AlN due to the high reactivity of both N and Ti.

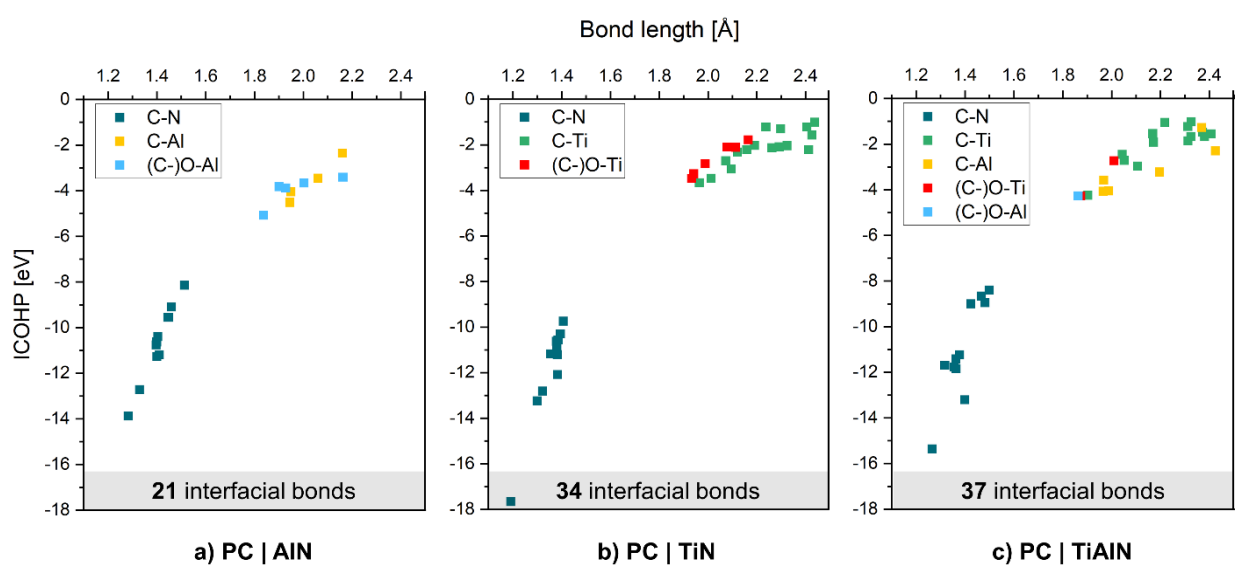

**Figure 7.** ICOHP analysis of individual interfacial bonds at the simulated a) PC | AlN, b) PC | TiN, as well as, c) PC | TiAlN interfaces and their total number of interfacial bonds.

### 3.4. Interfacial bond density and *indicator for adhesion*

To compare different interfaces with respect to the interfacial bond density, as well as the bond strength at the interface, Ref.[31] defined an *indicator for adhesion* for the metallic PC | Y interfaces (Y = Ti, Al, TiAl) by multiplying the calculated absolute ICOHP values (bond strength) of interfacial bonds with the measured bond density. In this way, the strength of different interfaces can be compared as a function of the interfacial bond formation (other factors such as interfacial stress and impurities are neglected in this analysis).

As a measure for the interfacial bond density, the area fraction of interfacial groups relative to the whole C 1s signal is determined based on XPS data for thin films with a comparable thickness (formula for *C 1s interfacial component fraction* see **Figure 8**). Due to the overlapping BE of (C-O)-metal and C-N groups in the C 1s signal, the N-C components of the N 1s spectra are also taken into account by considering the respective C 1s and N 1s sensitivity factors (more details can be found in the supporting information, **Table S4**).

As summarized in **Figure 8**, the highest density of interfacial bonds is obtained for PC | TiAlN, which is in good agreement with the simulations (**Figure 7**) and explainable by the complex cross-linking of interfacial groups. Experimentally, the PC | AlN and PC | TiN interfaces exhibit a similar overall density of interfacial groups, but differences in the composition are evident (**Figure 8**): For PC | AlN, most interfacial groups are due to the formation of C-N bonds (~ 8%), whereas (C-O)-Al and C-Al bonds (~ 2%) play only a minor role in the interface formation. For PC | TiN, however, the density of C-N groups (~ 6%) compared to the combined density of C-Ti and (C-O)-Ti groups (~ 4%) exhibit a lower difference compared to PC | AlN, indicating a high reactivity of both Ti and N atoms at the interface with PC.

When the interfacial bond density of the metal nitride thin films is compared to the metallic system based on Ref.[31], Al and TiAl thin films exhibit a significantly lower amount of interfacial bonds compared to the metal nitride systems, whereas Ti exhibits a similar interfacial bond density compared to the PC | TiAlN interface. It appears that the competing reactivity of Ti and N at the PC | TiN interface results in an overall smaller interfacial bond density compared to the PC | Ti interface (**Figure 8**). In contrast, the presence of N is preferential for the sputter deposition of both

PC | AlN and PC | TiAlN interfaces due to a significantly more frequent bond formation at the interface (**Figure 8**).

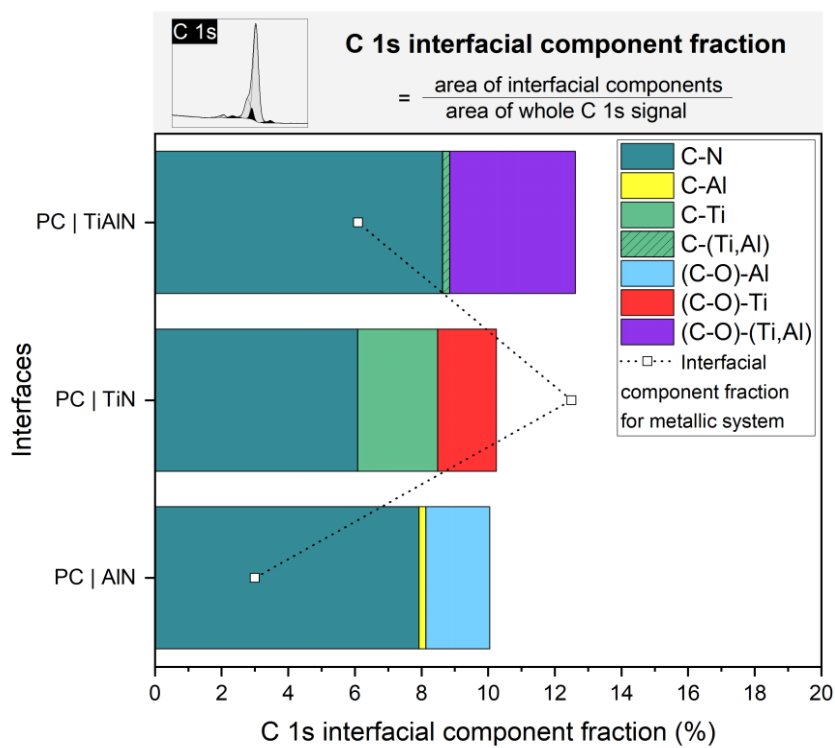

**Figure 8.** *C 1s interfacial component fraction* of different interfacial groups is shown for PC | X interfaces (X = AlN, TiN, TiAlN) and compared to the interfacial fraction obtained for the respective metallic PC | Y interfaces (Y = Al, Ti, TiAl) determined in Ref.[31] (dotted lines between data points are included as a guide for the eye).

To obtain in the next step the *indicator for adhesion*[31] the average absolute ICOHP value of interfacial bonds (**Figure 7**, **Table S5**) is multiplied by the respective *C 1s interfacial component fraction* (**Figure 8**). The *indicator for adhesion* with respect to the PC | X interfaces (X = AlN, TiN, TiAlN) is shown in **Figure 9** and supplemented by the values determined previously for the metallic interfaces.[31] Based on **Figure 9**, PC | TiAlN is identified as the strongest interface, followed by PC | AlN, while PC | TiN is the weakest. Comparing the contribution of different bond types to the overall adhesion indicator, the C-N bonds exhibit by far the highest share (**Figure 9**). In this way, all three metal nitride systems form stronger interfaces compared to the metal systems (black data points). **Figure 9** shows that the adhesion trend for the metal nitride interfaces differs from the metallic system since the PC | Al interface was identified as the weakest metallic interface,[31] while AlN forms a stronger interface compared to TiN. This can be rationalized by the low reactivity of Al atoms that is observed for the metallic as well as for the metal nitride interfaces with PC. While this low reactivity is unfavorable for the PC | Al interface due to the resulting low interfacial bond density (**Figure 8**), it seems to be favorable for the PC | AlN interface due to the formation of many strong interfacial C-N bonds. In contrast, the high reactivity of Ti atoms at the PC | TiN interface seems to be a disadvantage since both Ti and N tend to form many bonds at the interface and thus mutually compete. Consequently, fewer strong C-N bonds are formed compared to the PC | AlN interface and many weak C-Ti and (C-O)-Ti bonds are formed instead (**Figure 8**). For the PC | TiAlN interface, also C-N bonds play a major role in the interfacial bond formation (**Figure 7**). Additionally, the combination of highly reactive N and Ti, as well as selectively reactive Al, leads to a complex cross-linking of interfacial groups (instead of cluster formation) (**Figure 2**), resulting in the highest interfacial bond density observed theoretically (**Figure 7**) as well as experimentally (**Figure 8**). This interplay of many interfacial C-N bonds as well as a strong cross-linking of interfacial groups seems to promote the stronger interface formation for PC | TiAlN compared to the PC | AlN and PC | TiN interfaces.

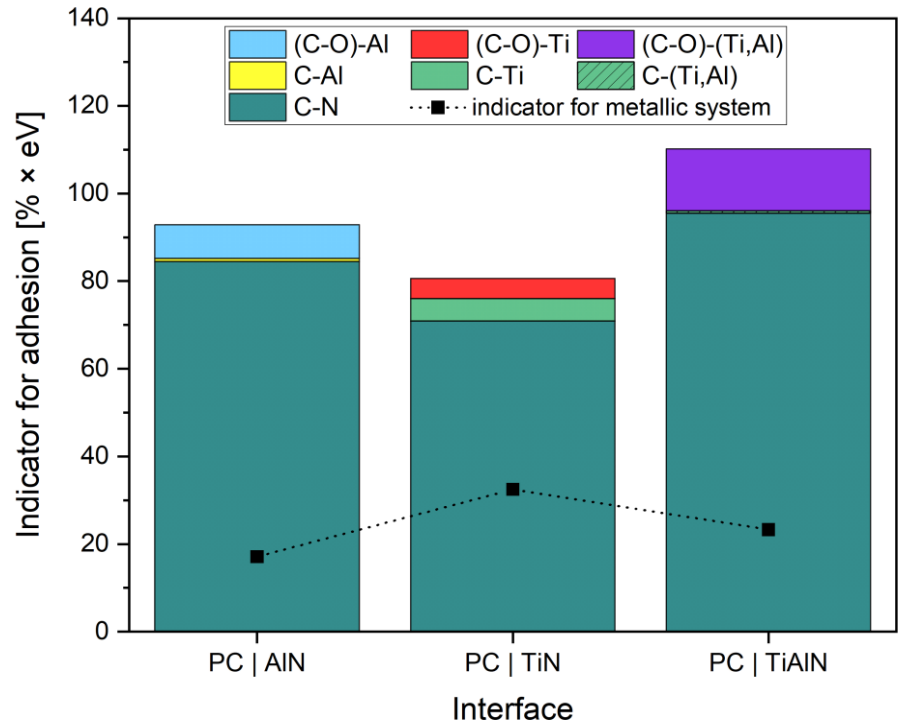

**Figure 9.** Combination of the measured relative interfacial bond density and the absolute ICOHP values of the respective bond type as an *indicator for adhesion* for the PC | X interfaces (X = AlN, TiN, TiAlN). The overall values for the adhesion indicator of the metallic PC | Y interfaces (Y = Al, Ti, TiAl) are indicated by the black data points[31] (dotted lines between data points are included as a guide for the eye).

## 4. Conclusion:

In this study, a 396 atoms-containing periodic structural model of PC was developed and subsequently bombarded by several Ti, Al, and N atoms to simulate interfacial bond formation during vapor deposition of AlN, TiN, and TiAlN onto PC using *ab initio* molecular dynamics. The simulations predict a high reactivity towards all functional groups of PC for both Ti and N atoms, whereas Al atoms selectively react with the carbonate group of pristine PC. However, after multiple bombardment events, the formation of activated surface groups enables bond formation between Al and C-N groups as well as some C radicals. Hence, Ti and N are predicted to contribute equally to the bond formation at the PC | TiN interface, whereas the PC | AlN interface is mostly defined by C-N groups, with Al-rich clusters forming around these groups. XPS data of the corresponding synthesized interfaces show good agreement with the predictions as the formation of C-N, C-Ti, C-Al, and (C-O)-(Ti,Al,N) bonds is experimentally verified. When combining the calculated bond strength (ICOHP) of the interfacial groups with the experimentally detected interfacial bond density as an *indicator for adhesion*, PC | TiAlN is identified as the strongest interface, followed by PC | AlN, while PC | TiN exhibits the lowest value. This can be rationalized by the selective reactivity of Al atoms causing the formation of a high interfacial density of strong C-N bonds, whereas the high reactivity of Ti leads to many weak C-Ti and (C-O)-Ti bonds but fewer strong C-N bonds. For PC | TiAlN, a combination of many strong C-N bonds and a complex cross-linking of interfacial groups enables the highest density of interfacial bonds as well as the formation of the strongest interface. Overall, the presented predictions show good agreement with the experimentally determined interfacial bond formation during magnetron sputtering of metal nitrides onto PC. It is expected that this research strategy is readily adaptable to probe bond formation across other organic | inorganic interfaces.

## 5. Methods

### 5.1. Experimental details

For the substrate preparation, 5 wt.% PC pellets (additive-free, product # 4315139, Sigma-Aldrich) were dissolved in tetrahydrofuran (anhydrous, 99.9% purity, inhibitor-free, Sigma-Aldrich) and spin-coated with a velocity of 2000 rpm on $1 \times 1$ cm$^2$ fused-silica substrates (Siegert Wafer GmbH). Afterwards, the prepared substrates were immediately inserted into the deposition chamber to minimize surface contaminations. Besides PC substrates, conductive $1 \times 1$ cm$^2$ Si(001) substrates (Crystal GmbH) were used for the deposition of thin films with a thickness > 100 nm to analyze the chemical composition.

All depositions were carried out in a lab-scale deposition chamber by using reactive direct current magnetron sputtering (DCMS). The Al (99.99% purity), Ti (99.995% purity), and composite $Ti_{0.5}Al_{0.5}$ (99.99% purity) targets (Ø 50 mm) were mounted at a target-to-substrate distance of 10 cm and an angle of 45° between target and substrate normal. The base pressure before all depositions was $< 6 \times 10^{-5}$ Pa. During sputtering, the substrate holder was kept electrically floating, while a rotation of 28 rpm was applied, enabling a homogenous deposition. No intentional heating was applied to the substrate. For reactive sputtering, an Ar partial pressure of 0.53 Pa was used, while the N$_2$ partial pressure was optimized to 0.13, 0.05, and 0.07 Pa (both gases ≥ 99.999% purity) to obtain close to the aimed stoichiometric $Al_{0.5}N_{0.5}$, $Ti_{0.5}N_{0.5}$, and $Ti_{0.25}Al_{0.25}N_{0.5}$ thin films, respectively. The power density applied to all three targets was kept constant at 10.2 W cm$^{-2}$. Before each deposition, the sputtering process at the target was already running for > 1 min behind a closed shutter to stabilize the process and clean the target surface of impurities.

Metal nitride samples deposited onto Si substrates for ≥ 10 min were analyzed by energy dispersive X-ray spectroscopy (EDX) and scanning electron microscopy to determine the composition and the film thickness, respectively. For the chemical analysis of the AlN and TiN thin films, a TM4000Plus Tabletop scanning electron microscope (SEM) (Hitachi Ltd.) equipped with a Quantax75 detector was used, while a JEOL JSM-6480 SEM with an EDAX Genesis 2000 device was used to measure the composition of the TiAlN thin film. For all three material systems, an acceleration voltage of 10 kV was applied. Due to the overlapping Ti Lα and N Kα signals, elastic-recoil detection analyzed TiN[41] and TiAlN[42] standards were used for the chemical quantification of the respective thin films.

To determine the film thickness, the cross sections of the thin films were examined by using an FEI Helios Nanolab 660 equipped with a field-emission microscope. During the SEM image acquisition, the acceleration voltage and current were set to 10 kV and 50 pA, respectively.

For the interfacial bond analysis, the samples were inserted into an XPS AXIS Supra instrument (Kratos Analytical Ltd.) immediately after deposition (atmosphere exposure time < 5 min). The base pressure in the XPS analysis chamber was always $< 5 \times 10^{-6}$ Pa. For the acquisition of the survey scans, a pass energy of 160 eV and a step size of 0.25 eV were used (5 sweeps, dwell time of 100 ms), while the C 1s and N 1s high-resolution scans were performed alternatively with a pass energy of 20 eV and a step size of 0.05 eV (10 sweeps, dwell time of 100 ms). During acquisition, a charge neutralization (low-energy, electron-only source) was applied to compensate for any charging effects. For the analysis of the interface samples, the BE scale was calibrated with respect to the C 1s signal of the hydrocarbon group of PC at BE = 284.6 eV,[28] whereas the metal nitride thin films were calibrated with respect to the Fermi edge.[43] The binding energy scale of the spectrometer was calibrated using a sputter-cleaned Ag foil (Ag $3d_{5/2}$ at 368.2 eV). For XPS data fitting, the CasaXPS software package was used (Casa Software Ltd.) and a Shirley background[44] was subtracted for all spectra. For the quantitative analysis of the survey scans, the sensitivity factors of the manufacturer of the instrument were applied.[45] The chemical state of elements in the C 1s signal was investigated by fitting components with a Gaussian-Lorentzian (70%-30%) line shape and the full width at half maximum (FWHM) was constrained to be ≤ 1.5 eV. The different line shapes for the N 1s components can be found in the supporting information (**Table S1**-**Table S3**).

**5.2. Computational details**

Models of growing AlN, TiN, and TiAlN films on PC surface were prepared using *ab initio* molecular dynamics (AIMD) utilizing the OpenMX density functional theory (DFT) package[46-49] with the GGA-based exchange-correlation potential by Perdew, Burke, and Ernzerhof[50] and DFT-D3 dispersion correction.[51] Initially, three perpendicular polycarbonate chains with 396 atoms in total were placed in a cubic box, with periodic boundary conditions applied resulting in three infinite PC chains. This box was slowly compressed in an AIMD run by reducing the box side by 1‰ every 5 fs while keeping a constant temperature of 450 K using a Nosé–Hoover thermostat.[52] This was done until the mass density of the cell was equal to 1.21 g cm$^{-3}$, the reported experimental mass density of PC (for illustration see **Figure S1** in supporting information). The final structure was relaxed at 0 K and a vacuum slab was inserted at a plane, where it resulted in the least amount of disjoined bonds. The dangling bonds created in the process were then terminated with H. An AIMD run followed with a thermalization at 300 K and the thermalized structure was used as a starting point for three atomic-bombardment AIMD simulations with the surface being alternately bombarded by N and Al atoms for AlN (N, Ti for TiN and N, Ti, N, Al for TiAlN) with a kinetic energy of 1 eV (typical kinetic energies of sputtered species during DCMS are in the order of a few eV[39]). The time interval between the individual bombardments was 0.5 ps, and the MD run was conducted until 30 atoms were deposited, while the temperature was kept constant at 300 K. The final structure was again relaxed and N 1*s* and C 1*s* core electron binding energies (BEs) were calculated using the core-hole approach.[53]

Numerical atomic basis sets (H6.0-s2p1, C6.0-s2p2d1, O6.0-s2p2d1, N6.0-s2p2d1, Ti7.0-s3p2d1, and Al7.0-s2p2d1) were used for the AIMD and structural relaxations, while an enhanced basis set was used for the BE calculations (H6.0-s2p1, C7.0_1s-s4p3d2, O7.0_1s-s4p3d2,

N7.0_1s-s4p3d2, Ti7.0-s3p2d1, Al7.0-s2p2d1). A cutoff energy of 220 Ry was used for the AIMD compression, 300 Ry for the relaxations and AIMD atomic bombardment, and 400 Ry for BE calculations. All calculations were done at the gamma point only, with an MD timestep of 0.5 fs and the relaxation stopping criterion of $1 \times 10^{-6}$ Ry/Bohr. All calculational data with further details are available in the NOMAD Archive.[32]

To estimate the interatomic bonding characteristics and strengths of individual bonds between PC and metal nitride atoms, the crystal orbital Hamilton population analysis technique[54] was employed. The final structures of the AIMD simulation were taken as input structures and a section containing the PC structural unit as well as a region of the metal nitride bulk was excised at a distance of 15 Å from the basal plane. This reduction in system size ensured an acceptable computational cost, particularly in terms of required memory. A static DFT simulation using the Vienna *ab initio* Simulation Package (VASP, version 5.4.4, University of Vienna)[55–57] was then performed to obtain the wave function of the system. Projector-augmented waves (PAW)[58,59] with a cut-off energy of 500 eV were used for basis set representation with electronic exchange and correlation effects accounted for using the generalized gradient approximation-based PBE functional.[50] The tetrahedron method by Blöchl[60] was employed for Brillouin zone integration over a $2 \times 2 \times 2$ *k*-mesh constructed using the Monkhorst-Pack approach.[61] The valence electron configurations specified in the atomic PAW potentials were $4s^13d^3$ for Ti, $3s^23p^1$ for Al, $2s^22p^3$ for N (atomic species in the metal nitrides), $2s^22p^2$ for C, $2s^22p^4$ for O, and $1s^1$ for H (atomic species in PC).

The obtained delocalized wave function of the system was then post-processed with the LOBSTER package (version 4.0.0, Institute of Inorganic Chemistry, RWTH Aachen University)[62–64] to project the delocalized wave function constructed from the PAW basis sets onto the individual atomic positions using the in-built local orbital basis set *pbevaspfit2015*.[65] Thus, atom-resolved orbital information was reconstructed, which allowed for crystal orbital Hamilton population (COHP) analyses to describe interatomic orbital mixing, and thus, the bonding characteristics. The integrated COHP values (ICOHP) of individual bonds were then taken as descriptors of the bond strength; while not identical, both properties are strongly correlated and the ICOHP is routinely used in literature to estimate bond strengths.[66–69]

**Acknowledgments**

This research was funded by the German Research Foundation (DFG, SFB-TR 87/3) "Pulsed high power plasmas for the synthesis of nanostructured functional layers" and project LM2023039 funded by the Ministry of Education, Youth and Sports of the Czech Republic. The authors gratefully acknowledge the computing time granted by the JARA Vergabegremium and provided on the JARA Partition part of the supercomputer CLAIX at RWTH Aachen University (JARA-HPC, project JARA0151, and JARA0221).

**CRediT author statement**

**Lena Patterer:** Conceptualization, Methodology, Formal analysis, Investigation, Writing – original draft, Visualization. **Pavel Ondračka:** Methodology, Formal analysis, Investigation, Writing – original draft. **Dimitri Bogdanovski:** Methodology, Formal analysis, Investigation, Writing – original draft. **Stanislav Mráz:** Methodology, Investigation, Writing – review & editing. **Soheil Karimi Aghda:** Formal analysis, Investigation, Writing – review & editing. **Peter J. Pöllmann:** Formal analysis, Investigation, Writing – review & editing. **Yu-Ping Chien:** Formal analysis, Investigation, Writing – review & editing. **Jochen M. Schneider:** Conceptualization, Methodology, Writing – original draft, Supervision, Project administration, Funding acquisition.

# Supporting information

## Correlative Theoretical and Experimental Study of the Polycarbonate | X Interfacial Bond Formation (X = AlN, TiN, TiAlN) during Magnetron Sputtering

*Lena Patterer[1]\*, Pavel Ondračka[2], Dimitri Bogdanovski[1], Stanislav Mráz[1],*

*Soheil Karimi Aghda[1], Peter J. Pöllmann[1], Yu-Ping Chien[1], Jochen M. Schneider[1]*

[1] Materials Chemistry, RWTH Aachen University, Kopernikusstr. 10, 52074 Aachen, Germany

[2] Department of Physical Electronics, Faculty of Science, Masaryk University, Kotlářská 2, 611 37 Brno, Czech Republic

\*Corresponding author: patterer@mch.rwth-aachen.de

1. Line shapes for N 1s components

**Table S1.** Line shapes of N 1s components fitted for an AlN thin film and the PC | AlN interface.

|          | (N,O)-Al      | C-N-Al  |
|----------|---------------|---------|
| **AlN**      | GL(70)T(1.1)  |         |
| **PC \| AlN** | GL(70)T(1.1)  | GL(30)  |

**Table S2.** Line shapes of N 1s components fitted for a TiN thin film and the PC | AlN interface.

|              | Ti-N       | (N,O)-Ti   | satellite  | C-N-Ti  | C-N    |
|--------------|------------|------------|------------|---------|--------|
| **TiN**      | GL(90)[1]  | GL(0)[1]   | GL(0)[1]   |         |        |
| **PC \| TiN** | GL(90)[1]  | GL(0)[1]   | GL(0)[1]   | GL(0)   | GL(30) |

**Table S3.** Line shapes of N 1s components fitted for a TiAlN thin film and the PC | AlN interface.

|                | N-(Ti,A)  | (N,O)-(Ti,Al) | satellite | C-N-(Ti,Al) |
|----------------|-----------|---------------|-----------|-------------|
| **TiAlN**      | GL(30)    | GL(30)        | GL(0)     |             |
| **PC \| TiAlN** |           | GL(30)        |           | GL(30)      |

2. Simulated PC bulk model

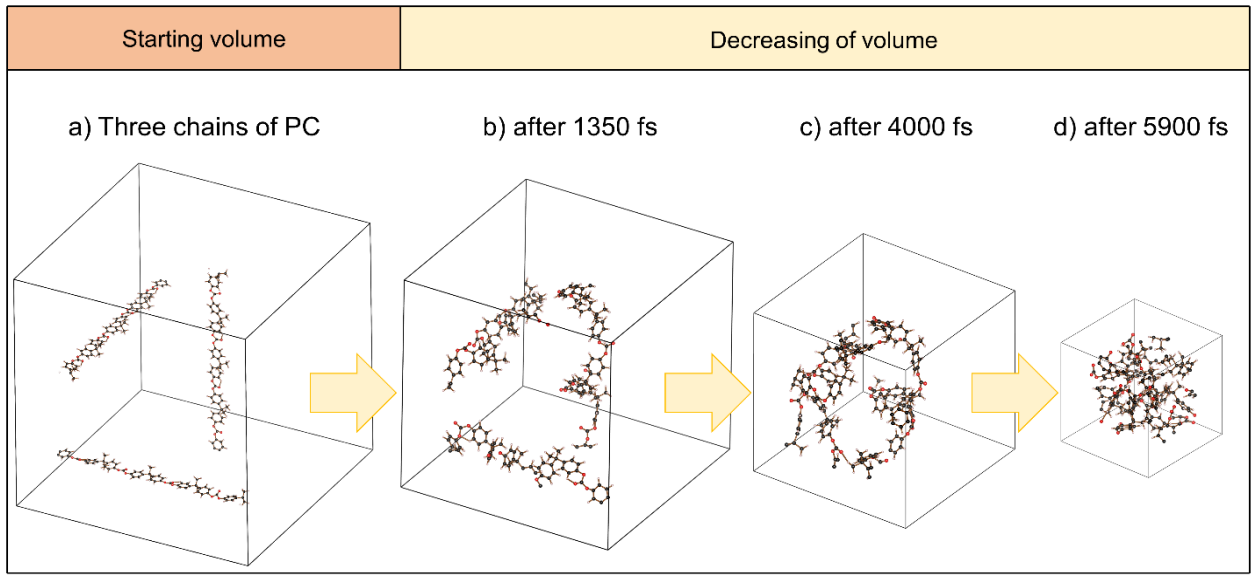

**Figure S1**. Schematic of the PC bulk model preparation visualized by VESTA.[2]

3. XPS N 1s spectra of two PC | TiN interfaces with different thicknesses

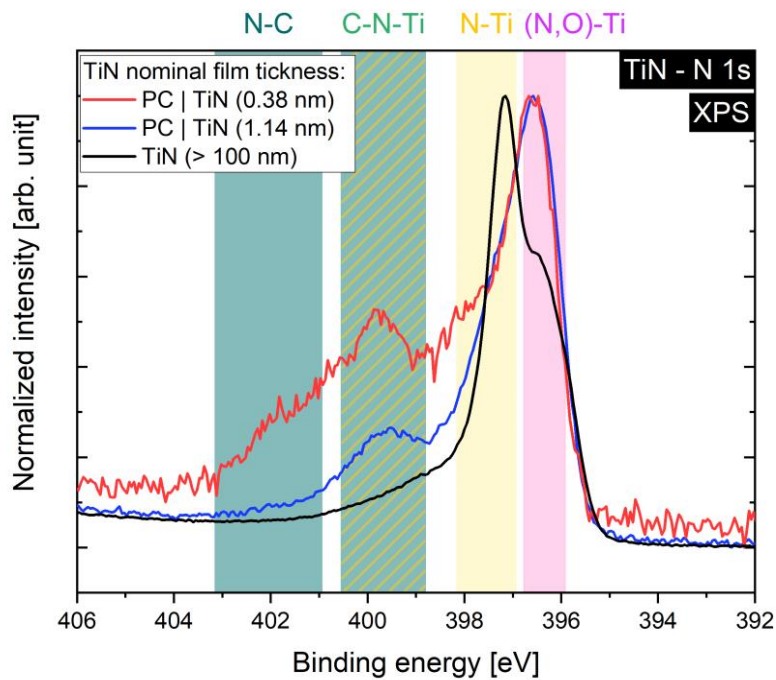

**Figure S2.** XPS N 1s spectra of two PC | TiN interfaces with different thicknesses and a TiN thin film (thickness > 100 nm). The initial N-C component reacts with Ti to form C-N-Ti during film growth (see red and blue curve).

### 4. Calculation of the *C1s component fraction* and the *indicator for adhesion*

**Table S4.** The overall concentrations of the interfacial groups are calculated from the XPS spectra (survey, N 1s, C 1s). Since the C-N + (C-O)-metal components are overlapping in the C 1s spectrum, the N-C groups of the N 1s spectrum are considered as well by calculating the overall concentration from the survey scan and subtracting it from the overall concentration of the C-N + (C-O)-metal components determined from the C 1s and survey spectra. The maximum concentration of C-N groups is constrained to the overall concentration of the C-N + (C-O)-metal components in the C 1s spectrum (if (2) > (4), then (5) = (3); this is the case for the PC | AlN and PC | TiN interfaces). For the PC | TiAlN interface, the concentration of N-C groups (2) is smaller compared to the concentration of the C-N + (C-O)-metal groups (4). The remaining fraction of the C-N + (C-O)-metal groups is added to the concentration of (C-O)-metal groups (see line (6)).

| Interfacial groups | Area fractions | PC \| AlN | PC \| TiN | PC \| TiAlN | |
|---|---|---|---|---|---|
| N-C | fraction in N 1s [%] | 68.81 | 21.88 | 87.22 | (1) |
| | fraction in survey [at.%] | 4.54 | 3.10 | 3.97 | (2) |
| C-N + (C-O)-metal | fraction in C 1s [%] | 7.93 | 6.09 | 11.81 | (3) |
| | fraction in survey [at.%] | 3.76 | 2.67 | 5.43 | (4) |
| C-N | fraction in C 1s [%] | 7.93 | 6.09 | 8.63 | (5) |
| (C-O)-metal | fraction in C 1s [%] | 1.92 + 0 | 1.76 + 0 | 0.59 + 3.18 | (6) |
| C-metal | fraction in C 1s [%] | 0.20 | 2.40 | 0.22 | (7) |

**Table S5.** Average absolute ICOHP (including the statistical uncertainty as standard deviation) determined from data points in **Figure 7**[3] and the resulting *indicator for adhesion* for different interfacial groups:

Adhesion indicator = 'average absolute ICOHP' × 'C 1s interfacial component fraction'
(values for the *C 1s interfacial component fraction* are color-marked in **Table S4**).

| | C-N | | (C-O)-metal | | C-metal | |
|---|---|---|---|---|---|---|
| | Average absolute ICOHP [eV] | Adhesion indicator [% × eV] | Average absolute ICOHP [eV] | Adhesion indicator [% × eV] | Average absolute ICOHP [eV] | Adhesion indicator [% × eV] |
| PC \| AlN | 10.65 ± 1.55 | 84.46 ± 12.29 | 3.97 ± 0.58 | 7.63 ± 1.57 | 3.77 ± 0.80 | 0.75 ± 0.23 |
| PC \| TiN | 11.64 ± 2.12 | 70.90 ± 12.91 | 2.59 ± 0.64 | 4.56 ± 1.59 | 2.14 ± 0.76 | 5.14 ±.2.58 |
| PC \| TiAlN | 11.06 ± 1.89. | 95.53 ± 16.31 | 3.75 ± 0.73. | 14.13 ± 3.89 | 2.29 ± 1.03 | 0.50 ± 0.32 |